# Electronic properties of zigzag graphene nanoribbons on Si(001)

Zhuhua Zhang[1] and Wanlin Guo[1, 2 a)]

[1]*Institute of Nano Science, Nanjing University of Aeronautics and Astronautics,*

*Nanjing 210016, China*

[2]*Department of Physics and High Pressure Science and Engineering Center,*

*University of Nevada, Las Vegas, Nevada 89154, USA*

We show by first-principles calculations that the electronic properties of zigzag graphene nanoribbons (Z-GNRs) adsorbed on Si(001) substrate strongly depend on ribbon width and adsorption orientation. Only narrow Z-GNRs with even rows of zigzag chains across their width adsorbed perpendicularly to the Si dimer rows possess an energy gap, while wider Z-GNRs are metallic due to width-dependent interface hybridization. The Z-GNRs can be metastably adsorbed parallel to the Si dimer rows, but show uniform metallic nature independent of ribbon width due to adsorption induced dangling-bond states on the Si surface.

[a)] Electronic mail: wlguo@nuaa.edu.cn

Graphene exhibits many outstanding physical properties, such as high electron mobility and long coherence length, due to a linear, neutrinolike energy spectrum near the Dirac point.[1] However, a major obstacle for graphene-based electronics is the inability to electrostatically confine electrons, which is naturally zero-gap semiconductor.[2] Two essential routes have been developed to induce a gap in the band structure of graphene. A straightforward solution is to pattern the graphene sheet into graphene nanoribbons (GNRs).[3-5] A confinement-induced gap can open in the GNRs owing to the presence of edges, which can be functionalized and shaped in various ways to achieve colorful properties.[6-8] Especially, the zigzag-edged GNR (Z-GNR) is predicted to exhibit an antiferromagnetic insulating ground state,[5,9] giving rise to unusual behaviors such as half metal[6,10] and giant magnetoresistance[11] and promising applications. Another route to open the gap is more practical to break the symmetry between sublattice in graphene by setting it onto a solid substrate or in applied fields.[12-15] A single-layer graphene on most of the substrates becomes a trivial buffer layer due to the breaking of the hexagonal π-orbital network as a result of strong bonding to the supports,[16-18] and a second layer is often necessary to be functional on a substrate. Although these routes are helpful for certain applications, none of them consider the essential interaction between the graphene edge and the substrate. To form meaningful quasi-one-dimensional nanodevices, use of GNRs on realistic substrates should be a more promising route. However, how a graphene with finite width adsorbed on substrate to form a functional system remains completely elusive. Here, we show by first-principles calculations that a single-layer Z-GNR adsorbed on Si(001) substrate can act as a functional layer with electronic properties sensitively dependent on ribbon width and adsorption orientation. These findings provide a useful guide for directing the construction of graphene-based devices on Si substrates.

All the calculations are performed within the framework of density functional theory as implemented in VASP code.[19] Ultrasoft pseudopotentials for the core region and the generalized gradient approximation for the exchange-correlation potential are used.[20] A kinetic energy cutoff of 530 eV is used in the plane-wave expansion. Following standard convention,[10] we denote a Z-GNR containing $N$ zigzag atomic chains across its width as $Z_N$-GNR. We vary the ribbon size from $Z_2$- to $Z_{14}$-GNRs, with widths from 2.74 Å to 27.97 Å. The surface model consists of eight Si monolayers in a 2×6 (or 2×8 and 2×10) surface unit cell, with two adjacent image surfaces separated by a vacuum region of 14 Å. The dangling bonds of Si atoms of the bottom monolayer and C atoms at the edges of GNRs are uniformly terminated by H atoms. The positions of the topmost six monolayer Si atoms plus the entire GNR atoms are fully relaxed until the force on each atom is less than 0.02 eV/Å. The 2D Brillouin-zone integration is sampled by up to 12 special $k$ points. Three primitive unit cells of Z-GNRs are chosen to match double Si(001) surface unit cell, with the GNRs stretched by about 3%.

To search all possible adsorption positions, the Z-GNRs are rigidly moved on the Si(001) surface by 5 Å per step along the directions parallel and perpendicular to the Si dimer rows, respectively. Then the structural optimization at each step is performed. We compared the binding energies[21] of all the optimized structures to find out the most stable adsorption configurations. As a result, for all the studied GNRs, the energetically most stable adsorption configuration is to bind perpendicularly to the Si dimer rows symmetrically at its edges with covalent Si-C bonds, rendering the GNR a bridge-like structure on the substrate, as shown in Figs. 1(a) and (b) for the $Z_6$- and $Z_7$-GNRs, respectively. This forms a striking contrast to graphene bonding to epitaxial substrates.[16,17] Such adsorption feature originates from the active edge states that are dominantly localized on the C atoms at ribbon edges. We will focus our discussions on the structural and electronic properties of these most favorable configurations. We typically discuss the structural properties of the adsorbed $Z_6$- and $Z_7$-GNRs shown in Fig. 1, which are labeled as $E_6$ and $E_7$, respectively. $E_6$ has a binding energy of 2.89 eV (per supercell), higher than 2.43 eV of $E_7$. In $E_6$, the energy cost by the deformation in the GNR and substrate are 1.09 eV and 3.34 eV, respectively; while for $E_7$ the corresponding values are only 1.05 eV and 3.12 eV, respectively. In contrast, the energy gain from the interfacial bonding is 7.32 eV in $E_6$, larger than 6.59 eV in $E_7$. Such difference in interfacial bonding is also well reflected in the Si-C bond lengths: 2.01~2.02 Å in $E_6$, shorter than 2.02~2.04 Å in $E_7$.

The electronic properties of the most favorable adsorption configurations show interesting change with the ribbon width. When the GNR width is less than 13 Å ($N<7$), the adsorbed $Z_N$-GNRs with even $N$ are semiconducting, while the ones with odd $N$ are metallic [Fig. 1(e)]. However, wider Z-GNRs in the studied range show uniform metallic nature on the substrate. The adsorbed Z-GNRs with the same electronic properties exhibit similar character in their band structures. We thus only present the band structures of the semiconducting $E_6$ and metallic $E_7$ as shown in Figs. 1(c) and (d), respectively. Here, the ribbon edge is in the $\Gamma$-$J$ direction. Both the GNRs become nonmagnetic upon the interaction with substrate. The shade regions are the projected energy bands of the adsorbed GNRs with all the substrate atoms removed and the four Si-C bonds replaced by C-H bonds. Figure 1(c) presents that $E_6$ has an indirect gap of about 0.18 eV along the ribbon edge, which is smaller than 0.29 eV of a freestanding $Z_6$-GNR. Nevertheless, the adsorbed $Z_6$-GNR is a nonmagnetic semiconductor with the gap increasable by strained Si technology (not shown), while a freestanding Z-GNR shows a gap only in an antiferromagnetic ground state that exists below the Neel temperature. Wave function analysis shows that the bands $S_{12}$ are from the surface states on Si dimers 1, 2 [see Fig. 1(a)] and bands $S_3$ are from surface states on the Si dimers uncovered by the GNR. Two mixed GNR-surface states, $M_1$ and $M_2$, originate from the interaction of the GNR state with the valence band states from Si dimers 1, 2. In contrast, in the $E_7$ case shown in Fig. 1(d), the GNR state interacts with the conduction band states from all remaining Si dimers on the substrate, resulting in two hybrid states $M_1$ and $M_2$ across the Fermi level.

To understand the width-dependent electronic properties of the adsorbed Z-GNRs, we calculated plane-averaged difference charge density $\Delta n(y) = n_{GNR+substrate}(y) - n_{GNR}(y) - n_{substrate}(y)$ (y is along the [001] direction), as shown in Figs. 2(b) and (d) for $E_6$ and $E_7$, respectively. The Friedel-like oscillations of the charge density are observed in the two cases as a result of the electron screening effect. It is shown that considerable charge in $E_7$ transfers from the GNR to the substrate, making the $Z_7$-GNR *p*-doped, as supported by the total negative extra electron density of -3.8 e/nm$^3$ in the space above the interface; whereas charge in $E_6$ mainly transfers to the interface region to form stronger Si-C bonds, and both the $Z_6$-GNR and substrate are nearly neutral. The stronger interface bonding in $E_6$ is also well reflected in the local density of states of the bonded Si and C atoms as shown in Fig. 2(e), where the electronic states for $E_6$ are located in deeper levels but for $E_7$ the electronic states distribute more close to the Fermi level. This significant difference can be further explained based on the frontier-orbital theory. We plot the charge contours of the dangling-bond (DB) states participating in the formation of Si-C bonds as shown in Figs. 2(a) and (c). It is noticed that the DB states show evident difference between the two underlying substrates. The DB characters of the substrate in $E_6$ are more *s*-like with most of their charge density participating in bond formation, while the counterparts in $E_7$ have more *p*-like character, see Figs. 2(b) and (d). It is known that the *s*-like surface DBs have lower energy and lie at the valence band top while the *p*-like surface DBs form the conduction band bottom in the Si(001) surface.[22] So the $Z_6$-GNR state interacts with occupied valence band states of the substrate, whereas the $Z_7$-GNR state interacts with unoccupied conduction band states, in accord with the band plots shown in Figs. 1(c) and (d). Through extensive analysis to other adsorbed Z-GNRs, we conclude that the adsorbed Z-GNR shows metallic nature as soon as the substrate provides a *p*-like DB to join into the interface hybridization, simply because the *p*-like DB leads to relatively weak interface bonding that enables the charge transfer from the GNR to the substrate.

In general, the electronic properties correlate closely with the atomic geometry. Nevertheless, the deformation of GNRs induced by adsorption is found to hardly affect their electronic properties. We therefore examine the change of surface atomic structures to further understand our results. A remarkable contrast between $E_6$ and $E_7$ is from the back bond angles of the dimers binding to the GNR as denoted by $α$ (four in total for each configuration) in Fig. 1. It is found that two larger back bond angles are only 111.4° in $E_6$, close to the ideal $sp^3$ hybridization value of 109.5°, which thus makes all the Si DB states more $s$-like; whereas the two larger back bond angles remain $sp^2$-like 120.5° in $E_7$, thus giving two $p$-like Si DB states for formation of Si-C bonds. However, these structural differences become gradually indistinguishable with increasing the GNR width, and $p$-like DB states always join in the interface hybridization in the adsorption of wider Z-GNRs, mainly due to the reduction in lateral mismatch strain between the GNR and substrate. Therefore, wide Z-GNRs show uniform metallic nature on the substrate.

When the ribbon edge is parallel to the Si dimer rows, we find several metastable adsorption positions but the system keeps metallic nature independent of ribbon width. In this adsorption orientation, the GNR can only bond to one Si atom of the dimer, leaving a DB state on the rest atom,[23] as circled in Figs. 3(a) and (b). These DB states cross the Fermi level and make the whole hybrid structures metallic. The most favorable atomic structures for the $Z_6$- and $Z_7$-GNRs along this adsorption orientation are shown in Figs. 3(a) and (b), respectively, together with their band structures. In the $Z_6$-GNR, four covalent Si-C bonds, 2.05~2.06 Å, and in the $Z_7$-GNR only two

Si-C bonds of 2.02 Å are still uniformly formed at ribbon edges. The binding energies for both the sites are 2.08 eV and 2.11 eV, respectively, distinctly lower than the cases perpendicular to the dimer rows. Here, the ribbon edge is along the $\Gamma$-$J'$ direction. For the adsorbed $Z_6$-GNR, the Fermi level crosses both a GNR state, G, and a surface band, $S_1$. The state $S_1$ is just from the DB states on the Si atoms 1 as noted in Fig. 3(a) and the state G becomes partially occupied due to the charge transfer. In contrary, for the $Z_7$-GNR, three bands cross the Fermi level, with two of them, $S_1$ and $S_2$, from the DB states and the other one, S, from the remaining surface states, and no electronic channel is found with the $Z_7$-GNR. Close observation on the configuration clearly shows that the $Z_7$-GNR only bonds with two $sp^3$-like bonded "up" Si atoms of two Si dimers.[23] So the GNR state, G, mixes with valence band states of the substrate and the charge transfer is suppressed from the GNR to the substrate. This result further supports the aforementioned hybridization analysis.

In summary, we have revealed by first-principles calculations the stable adsorption configurations for Z-GNRs on Si(001) and their electronic properties that sensitively depend on ribbon width and adsorption orientation. Our results may pave a path toward graphene-based devices by integrating the emerging nanoscale graphene systems with existing Si technology.

This work is supported by the 973 Program (2007CB936204), National NSF (10732040), Jiangsu Province NSF (BK2008042), the MOE (IRT0534) of China, and Jiangsu Province Innovation Project for Graduate Student (CX07B_064z) of China.

[21] The binding energy of a Z-GNR adsorbed on Si surface is calculated by the energy difference between the total energies of the combined system and the sum of the total energies of a pristine GNR and a clean Si surface in the same simulation box.

[22] K. Seino, W. G. Schmidt, and F. Bechstedt, Phys. Rev. Lett. **93**, 036101 (2004).

[23] J. W. Lyding, T. C. Shen, and J. S. Hubacek, Appl. Phys. Lett. **64**, 2010 (1994).

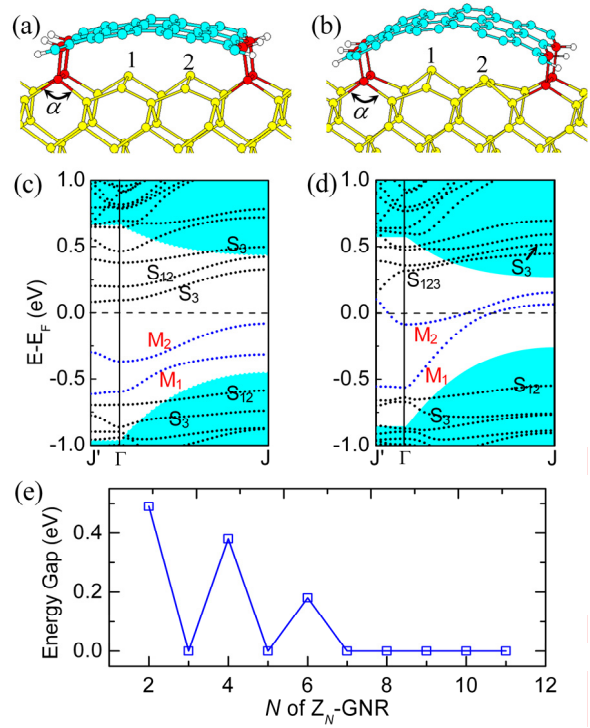

FIG. 1. (Color online) The most stable equilibrium configurations for the adsorbed (a) $Z_6$- and (b) $Z_7$-GNRs on Si(001). Yellow, cyan and white balls denote the Si, C and H atoms, respectively. Formed Si-C bonds in the adsorption are highlighted in red color. (c) and (d) show the corresponding band structures for the configurations illustrated in (a) and (b), respectively. The Fermi level is denoted by the dash line. The shaded regions display the projected energy bands of the adsorbed Z-GNRs with four C-Si bonds replaced by C-H bonds. (e) Energy gap of the adsorbed GNR system as a function of the number of zigzag atom chains $N$ in $Z_N$-GNRs.

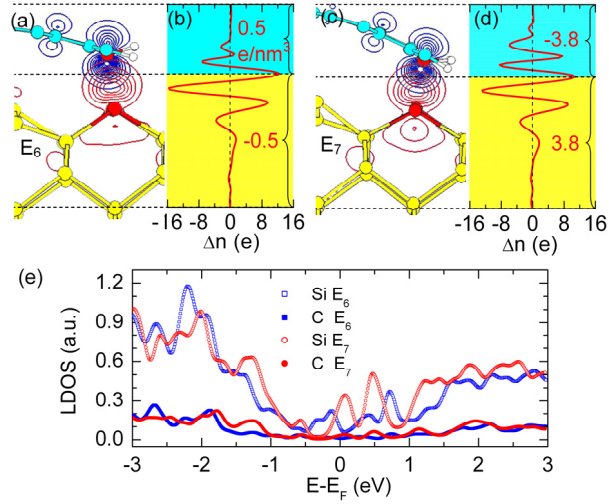

FIG. 2. (Color online) Interface hybridization analysis for Z-GNRs adsorbed on Si(001). (a), (c) Contour plots (spacing 0.85 e/Å$^3$) of dangling bonds states participating in the formation of covalent Si-C bonds. Red and blue lines denote the states on the GNR and Si substrate, respectively. (b), (d) Plane-averaged difference electron densities, with the longitudinal coordinate equal to the dimension along Si[001] direction, as denoted by the dash lines. The numbers denote the total extra electron densities in regions above (GNR) and below (substrate) the interface. (e) Local density of states for a pair of bonded Si and C atoms in $E_6$ and $E_7$.

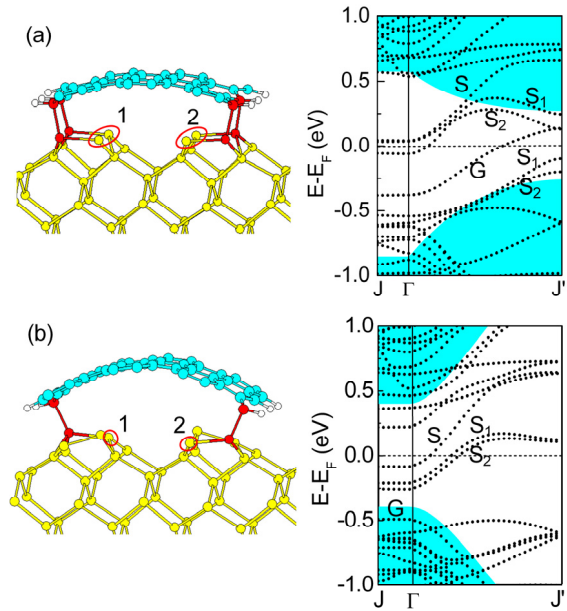

FIG. 3. (Color online) Atomic and electronic structures of the second most favorable configurations for (a) $Z_6$- and (b) $Z_7$-GNRs adsorbed on the Si(001) parallel to the Si dimer rows. States $S_1$ and $S_2$ are associated with dangling bond states on the atoms denoted by red circles, and state S is associated with the surface states on remaining Si dimers of the substrate. Other stipulations are the same as in Fig. 1.